\documentclass[%
reprint,
 amsmath,
 amssymb,
 aps,
 prl,
]{revtex4-2}

\usepackage[colorlinks=true,linkcolor=red,citecolor=blue]{hyperref}
\usepackage{graphicx}
\usepackage{dcolumn}
\usepackage{bm}

\usepackage[normalem]{ulem}
\newcommand{\stkout}[1]{\ifmmode\text{\sout{\ensuremath{#1}}}\else\sout{#1}\fi}

\begin{document}
\preprint{Manuscript}
\title{Dynamics of a Passive Droplet in Active Turbulence}
%
\author{Chamkor Singh}
\email{chamkor.singh@cup.edu.in}
\affiliation{Department of Physics, Central University of Punjab, Bathinda 151401, India}

\author{Abhishek Chaudhuri}
\email{abhishek@iisermohali.ac.in}
\affiliation{Department of Physical Sciences, Indian Institute of Science Education and Research Mohali, Manauli 140306, India}
\date{\today}
%
\begin{abstract}
We numerically study the effect of an active turbulent environment on a passive deformable droplet. The system is simulated using coupled hydrodynamic and nematodynamic equations for nematic liquid crystals with an active stress which is non-zero outside the droplet, and is zero inside. The droplet undergoes deformation fluctuations and its movement shows periods of ``runs" and ``stays". The mean square displacement of the geometric center of the droplet shows an extended ballistic regime and a transition to normal diffusive regime which depends on the size of the droplet. We relate this transition with a temporal scale associated with velocity autocorrelation function of the droplet trajectories, and with a spatial scale associated with one-time two-point velocity correlation function of the surrounding active medium. As the radius is decreased below the integral length scale, the velocity autocorrelation time of trajectories increases and the transition to normal diffusion is delayed.
\end{abstract}
%
%
\maketitle
%
%

The random motion of a passive tracer particle moving in a fluid at equilibrium is the starting point of describing fluctuating dynamics~\cite{chaikin1995principles}. Microrheology suggests that the local and bulk mechanical properties of a complex fluid can be extracted from the dynamics of such tracer particles~\cite{squires2010fluid,puertas2014microrheology}. A similar strategy can be used for an active fluid, where particle movement is coupled to nonequilibrium degrees of freedom. An active fluid consists of a collection of particles or cells suspended in a fluid, which can convert chemical energy into mechanical work by generating stresses at the microscale~\cite{lauga2009hydrodynamics,koch2011collective,marchetti2013hydrodynamics,saintillan2018rheology}. For instance, a flock in a fluid can be considered as a liquid crystal which is made of the swimming particles and an oriented state of the system possesses an `active stress' which is proportional to the orientation~\cite{ramaswamy2019active}. Active fluids have been shown to exhibit unusual transport properties such as enhanced diffusion, accumulation near boundaries, and rectification~\cite{elgeti2015physics,sokolov2012physical}. Experiments using passive particles in active fluids have revealed superdiffusive behavior of a passive particle at short times and a dramatically enhanced translational diffusion at long times~\cite{wu2000particle,kim2004enhanced,chen2007fluctuations,wilson2011differential,mino2011enhanced,jepson2013enhanced,valeriani2011colloids,leptos2009dynamics,kurtuldu2011enhancement,katuri2021inferring}.  

The collective dynamics of the suspended swimmers in the fluid such as microbial suspensions, cytoskeletal suspensions, self-propelled colloids, and cell tissues, generates spontaneous flows. These flows are characterized by chaotic spatiotemporal patterns and are referred to as active turbulence~\cite{alert2022active,thampi2016active}. Experimental studies of active turbulence look at the velocity fields and their statistics to characterize long range hydrodynamic flows ~\cite{wensink2012meso} and show the creation, transportation and annihilation of topological defects by active stresses~\cite{decamp2015orientational}. Theoretical studies of passive tracers in active fluids use the generalized Langevin equation 
with different properties of the friction and fluctuating force due to the fluid or use particle based simulations of passive and active objects~\cite{granek2022anomalous,zaid2011levy,maggi2014generalized,argun2016non,dabelow2019irreversibility,knevzevic2020effective,ye2020active,soni2003single,belan2021active,dunkel2010swimmer,zaid2011levy,foffano2012colloids,mino2013induced,abbaspour2021enhanced,ortlieb2019statistics,maitra2020enhanced,granek2022anomalous,angelani2011effective,maes2020fluctuating,peng2016diffusion,speck2021vorticity}. To the best of our knowledge, numerical studies of passive particle dynamics in active fluids have almost always neglected hydrodynamic interactions of the swimmers because of the complexity involved~\cite{kanazawa2020loopy} or have focused on the drag on the passive particle in the presence of applied force~\cite{foffano2012colloids}. Recent experiments of mobile {\em rigid} inclusions in an active nematic system have shown the importance of the complex interplay and feedback between the motion of such an inclusion and the active nematic~\cite{ray2023rectified}.  In this letter we computationally study the dynamics of a mobile {\em deformable} inclusion, which we call a droplet, in an active turbulent fluid.

\begin{figure*}[!t]
	\centering
	\centering
	\includegraphics[width=1.0\linewidth]{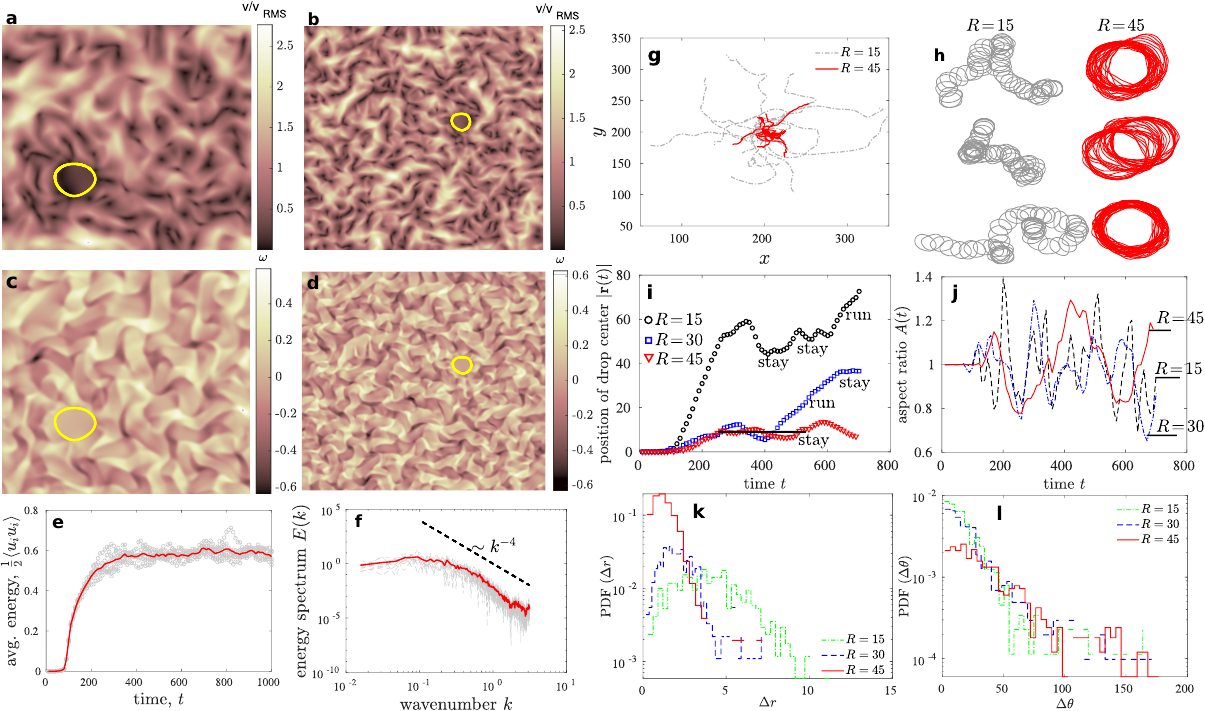}
	\caption{(a-b) Absolute velocity scaled by its root mean square value for system size $200\times200$ and $400\times400$ respectively. The droplet is shown with the solid yellow line. (c-d) Vorticity field for system size $200\times200$ and $400\times400$ respectively. (e) Average kinetic energy per unit mass with time. Gray symbols for individual realizations and the red line is ensemble average.  
	Time from $t=0$ to $t\approx 300$, when the active turbulence is still developing, is neglected and a time window of $300 \leq t \leq 800$ is considered for all the statistical results in this paper, including the above subfigures (a-d) which are taken at $t=500$. (f) Time averaged energy spectrum with spectral exponent close to $-4$, a behavior also observed by~\cite{giomi2015geometry,martinez2021scaling}.
	(g) Individual realizations of trajectories of the geometric center $\boldsymbol{r}(t)$ of passive nematic droplet of radius $R=15$ and $R=45$, and (h) droplet interface during typical realizations for $R=15$ and $R=45$. (i) Position of the droplet center relative to initial position, $|\boldsymbol{r}(t)|$ with time, exhibiting ``run" and ``stay" chracteristics along the droplet trajectories. (j) Ratio of the horizontal to vertical span of the droplet for $R=15$, $30$ and $45$. Dynamics of translations as well as the aspect ratio slows down as the radius is increased. (k) Probability distribution function of the translational steps taken by the geometric center of the droplet. PDF($\Delta r$) narrows down upon increasing the radius of the droplet. (l) PDF of discrete changes in the direction of motion of the droplet. PDF($\Delta \theta$) scales nearly exponentially for intermediate range of angles ($\Delta\theta\approx 25^o$ to $65^o$) with heavy tailed rare events for $>90^o$ or close to complete reversal.}
	\label{fig_energy_vorticity_velocity}
\end{figure*}

One of the theoretical approaches to study active turbulence is active liquid crystal theory where the continuum equations of motion are built from symmetry arguments. A particular class is active apolar nematic systems characterized by head-tail symmetry of the constituent particles -- tending to align with each other. These systems exhibit characteristics such as instability of the isotropic nematic state and intriguing pattern formation~\cite{thampi2015intrinsic,giomi2012banding,marchetti2013hydrodynamics,doostmohammadi2018active,alert2022active,colin2019chemotactic}, universal scalings of the energy spectra~\cite{alert2020universal,urzay2017multi,rorai2022coexistence}, topological chaos~\cite{tan2019topological}, and in general far from equilibrium dynamics~\cite{fodor2016far}. The homogeneous isotropic state is unstable and goes into a sequence of instabilities~\cite{martinez2019selection}. Once the system is in a developed state, the ensemble averaged mean kinetic energy in the system approches a statistically steady state~\cite{koch2021role} [Fig.~\ref{fig_energy_vorticity_velocity} (e)]. We suspend a passive nematic droplet at this stage of the active turbulence and study its dynamics [Fig.~\ref{fig_energy_vorticity_velocity} (a-d)]. Inverse to this setup are systems consisting of active droplets immersed in passive media which demonstrate complex behaviors, for example, spontaneous onset of motility and division of active nematic droplets~\cite{giomi2014spontaneous}, self-organization and division in active liquid droplets~\cite{weirich2019self}, dynamic defect structures in active nematic shells~\cite{zhang2016dynamic}, and emulsification in mixtures of active and passive component media~\cite{guillamat2018active}. 

We computationally characterize a system where the environment or the bath is an active liquid crystal and a passive nematic soft object is the suspended phase. It is known that an isotropic liquid droplet in a passive nematic liquid crystal under applied shear undergo different modes of movement (oscillatory, breakup, or motile) when the anchoring conditions at the surface of the droplet are changed~\cite{tiribocchi2016shear}. However inside an active turbulent medium, the forcing is far from simple shear, or random. The forces on the droplet are correlated over certain spatial and temporal scales and are an important aspect of its dynamics.

%
%
Apolar active systems in the continuum limit can be modeled using the passive nematic liquid crystal theory with added active stresses, combined with hydrodynamics. The orientation of nematics is described by a director field $\boldsymbol{n}$, with $\boldsymbol{n}\equiv-\boldsymbol{n}$ for apolar nematics. The symmetric traceless second rank tensor $\boldsymbol{Q} = q\frac{d}{d-1}(\boldsymbol{n}\boldsymbol{n}-\boldsymbol{I}/d)$ characterizes the nematic order, where $q$ is the strength of the nematic order, $\boldsymbol{I}$ is the identity tensor and $d$ denotes the spatial dimension. In our simulations, $d = 2$. The evolution of $\boldsymbol{Q}$ is described by the nematodynamic equation~\cite{doostmohammadi2018active,thampi2015intrinsic}
\begin{equation}
\partial_t\boldsymbol{Q} + \boldsymbol{u} \cdot \nabla \boldsymbol{Q}= \boldsymbol{S}+ \Gamma \boldsymbol{H}
\end{equation}
where
$
\boldsymbol{S}=(\lambda\boldsymbol{E}+\boldsymbol{\Omega})\cdot(\boldsymbol{Q}+\boldsymbol{I}/d) 
+(\boldsymbol{Q}+\boldsymbol{I}/d)\cdot(\lambda\boldsymbol{E}-\boldsymbol{\Omega})
-2\lambda(\boldsymbol{Q}+\boldsymbol{I}/d)(\boldsymbol{Q}:\nabla\boldsymbol{u})
$
is co-moving tensor with $\boldsymbol{E}=[\nabla\boldsymbol{u}+(\nabla\boldsymbol{u})^T]/2$ and $\boldsymbol{\Omega}=[\nabla\boldsymbol{u}-(\nabla\boldsymbol{u})^T]/2$ being the strain rate and vorticity tensors respectively. The tensor $\boldsymbol{H}$ describes relaxation of $\boldsymbol{Q}$ towards the minimum of a free energy and can be written as
$
\boldsymbol{H}=K\nabla^2\boldsymbol{Q}+C\boldsymbol{Q}/3+C(\boldsymbol{Q}\boldsymbol{Q}-\boldsymbol{Q}:\boldsymbol{Q}\;\boldsymbol{I}/d)-C\boldsymbol{Q}(\boldsymbol{Q}:\boldsymbol{Q})$, where $K$ is the elastic constant and $C$ is a material constant~\cite{thampi2015intrinsic}. The $\boldsymbol{Q}$-tensor field is coupled to velocity field $\boldsymbol{u}$ via the momentum equation
\begin{equation}
\rho\:(\partial_t\boldsymbol{u} + \boldsymbol{u} \cdot \nabla \boldsymbol{u})= \nabla\cdot(-p\boldsymbol{I}+2\eta\boldsymbol{E}+\boldsymbol{\sigma}^\textrm{P}+\boldsymbol{\sigma}^\textrm{A})-\mu\boldsymbol{u}+\boldsymbol{f}^\textrm{I},
\end{equation}
where the passive elastic stress tensor
$
\boldsymbol{\sigma}^\textrm{P}=2\lambda(\boldsymbol{Q}+\boldsymbol{I}/d)(\boldsymbol{Q}:\boldsymbol{H})
-\lambda\boldsymbol{H}\cdot(\boldsymbol{Q}+\boldsymbol{I}/d)
-\lambda(\boldsymbol{Q}+\boldsymbol{I}/d)\cdot\boldsymbol{H}
+\boldsymbol{Q}\cdot\boldsymbol{H}-\boldsymbol{H}\cdot\boldsymbol{Q}
-K\partial_iQ_{kl}\partial_jQ_{kl}
$~\cite{thampi2015intrinsic}. The stress tensor $
\boldsymbol{\sigma}^\textrm{A}=-\zeta\boldsymbol{Q}$ gives rise to activity or motility in the system~\cite{simha2002hydrodynamic}. The activity parameter $\zeta\geq0$ ($\leq0$)  for a contractile (extensile) active nematic. The term $-\mu\boldsymbol{u}$ mimics substrate friction or drag, and $p$ is the pressure. The fluid is considered incompressible so that $\nabla \cdot \boldsymbol{u} = 0$.

We suspend a droplet as a second phase inside the active medium. The force $\boldsymbol{f}^\textrm{I}$ is due to the interfacial tension and acts only at the interface between passive droplet and the outer active medium~\cite{tryggvason2011direct,unverdi1992front,singh2018levitation}. The coupled set of nematodynamic, momentum, and incompressibility equations are integrated
computationally, utilizing a finite-volume based pressure projection algorithm, utilizing the Eulerian one-fluid approach~\cite{singh2018levitation,tryggvason2011direct,bell1989second}. In this, the properties and parameters such as density, viscosity, activity, alignment, friction, etc. are also considered as fields and take different values in the two phases (see supplemental for parameter values in individual phases). The spatial distributions of the properties is updated with time using the front-tracking algorithm~\cite{tryggvason2011direct,unverdi1992front,singh2018levitation} in which the interface location is first updated using the velocity field solution and then the property fields and forces due to interfacial tension are reconstructed from the known interface location~\cite{tryggvason2011direct,unverdi1992front,singh2018levitation}. One feature of the present formulation is that the anchoring conditions at the surface of the suspended droplet emerge from the solution itself, instead of implementing them as an input. The advective fluxes in the continuum equations are reconstructed using a sixth order weighted essentially non-oscillatory scheme which is quite volume conservative~\cite{balsara2000monotonicity,titarev2004finite}. 

The system of equations is unstable around the state $\boldsymbol{Q}=\boldsymbol{0},\boldsymbol{u}=\boldsymbol{0}$; the system passes through states consisting of vortices and then proceeds towards spatiotemporal chaos. Eventually active stresses, on an average, are balanced by the passive, dissipative and frictional stresses and the system reaches a statistically steady state [Fig.~\ref{fig_energy_vorticity_velocity} (a-d)]. To characterize the active turbulent state, we look at the kinetic energy spectrum $E(k) = \langle \frac{1}{2\pi}\sum_{k\leq |\boldsymbol{k}|<k+\Delta k} |\hat{\boldsymbol{u}}(\boldsymbol{k},t)|^2 \rangle$ where $k$ is the magnitude of the wavevector or the wavenumber, $\hat{\boldsymbol{u}}(\boldsymbol{k},t)=\int \boldsymbol{u}(\boldsymbol{r},t)\exp(- i\boldsymbol{k}.\boldsymbol{r})d\boldsymbol{r}$, and $\langle \rangle$ denotes time average. We find a power law scaling $E(k)\sim k^{-4}$ [Fig.~\ref{fig_energy_vorticity_velocity} (f)] which has also been observed by~\cite{giomi2015geometry,martinez2021scaling}.

\begin{figure*}[!t]
	\centering
	\includegraphics[width=1.0\linewidth]{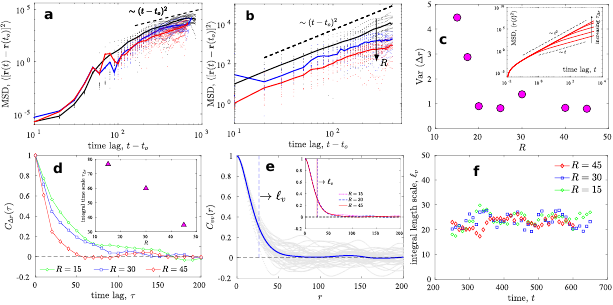}		
	\caption{(a) Mean square displacement of droplets of radius $R=15$, $30$ and $45$ with $t_o=0$, and (b) zoom into the super diffusive regime with $t_o=300$. (c) Variance of discrete step sizes taken by droplets of different radii. The variance decreases as the droplet radius increases; (c, inset) Mean square displacement shows crossover from ballistic to diffusive regime for a particle under exponentially correlated random force with time scale $\tau_F$. The crossover depends on $\tau_F$, or the velocity correlation time $\tau_{\Delta r} \sim m$. (d) Step autocorrelation function $C_{\Delta r}(\tau)$ and (d, inset) associated integral time scale $\tau_{\Delta r}\equiv \int_{0}^{\infty} C_{\Delta r}(\tau)d\tau$. Smaller droplets exhibit relatively larger $\tau_{\Delta r}$, or increased memory of the history of steps taken. (e) Time averaged one-time two-point velocity correlation function of the velocity field in active nematic environment. The integral length scale $\ell_v\equiv \int_{0}^{\infty} C_{vv}(r)dr$ is marked with a vertical dashed line; (e, inset) $\ell_v$ in the active bath changes negligibly upon changing the droplet radius. (f) Integral length scale with time during typical realizations.}
	\label{fig_corr_velocity_in_space}
\end{figure*}

Droplets of radii in the range $15\leq R \leq 45$ are suspended in the active turbulent state, and for a given radius the simulations are repeated at least $15$ times to obtain an ensemble of stochastic trajectories. A typical set of trajectories of the geometric center of the droplet is shown in Fig.~\ref{fig_energy_vorticity_velocity} (g-h) for radius $R=15$ and $R=45$. To be precise, the position of the center of the droplet is defined as $\boldsymbol{r}(t)=\sum_i \boldsymbol{x}_i(t)/N$ where the perimeter of the droplet is divided into $N$ points having positions $\boldsymbol{x}_i$ -- consistent with the front-tracking algorithm~\cite{tryggvason2011direct,unverdi1992front,singh2018levitation}. As the droplet interacts and is forced by the surrounding active bath, we measure its distance relative to the initial position with time $|\boldsymbol{r}(t)|$; typical cases are depicted in Fig.~\ref{fig_energy_vorticity_velocity} (i). Along the trajectories, the droplets undergo periods of ``runs" and ``stays" as marked on the plot. The ``stays" are prolonged as the radius of the droplet is increased. We also note that the time scale of undulations of droplet deformation parameter $A(t)$, defined as the ratio of horizontal to vertical span of the droplet, increases as we increase the radius [Fig.~\ref{fig_energy_vorticity_velocity} (j)]. With increasing size the droplet dynamics slows down and we transit from a ``fast" process towards a ``slow" process.

The trajectories consist of discrete translational steps taken by the droplet center $\Delta r\equiv |\boldsymbol{r}(t+\Delta t)-\boldsymbol{r}(t)|$ in time step $\Delta t$. In addition to translational steps $\Delta r$, we define angular steps $\Delta \theta=\cos^{-1}\left[ \boldsymbol{a}\cdot\boldsymbol{b}/ab\right]$ where $\boldsymbol{a}=\boldsymbol{r}(t+\Delta t)-\boldsymbol{r}(t)$ and $\boldsymbol{b}=\boldsymbol{r}(t)-\boldsymbol{r}(t-\Delta t)$. Distribution of these two quantities are shown in Fig.~\ref{fig_energy_vorticity_velocity} (k-l). The step length distribution (or equivalently the step velocity distribution) spreads upon decreasing the radius of the droplet [Fig.~\ref{fig_energy_vorticity_velocity} (k)]. Smaller droplets execute relatively larger translational steps. The change in direction of motion of the droplet is nearly exponentially distributed for intermediate range of angles ($\Delta\theta\approx 25^o$ to $65^o$), and exhibits some rare events of $>90^o$ or close to complete reversal of the direction [Fig.~\ref{fig_energy_vorticity_velocity} (h,l)]. Assuming a functional form for the translational distribution $f(\Delta r) = A\: e^{-\alpha (\Delta r)^2}$, the translational diffusion can be calculated as $D=\int_{-\infty}^{\infty}(\Delta r)^2f(\Delta r)d(\Delta r)/2\tau=\sqrt{\pi}A/(4\tau \alpha^{3/2})$ with discretization time step $\tau$. This implies a diffusive mean square displacement (MSD) $\sim Dt$ for $t\gg\tau$. However, the computation of MSD as  shows a superdiffusive regime where $\langle \Delta{\bf r}^2(t)\rangle \sim t^{\delta}$ with power exponent $1.5 \leq {\delta} \leq 2.0$, before any transition to  normal diffusion (see Fig.~\ref{fig_corr_velocity_in_space} (a-b)). This is consistent with the behaviour observed in experiments of tracer diffusion in bacterial turbulence.

%
To further understand the droplet dynamics, we compute the normalized step autocorrelation function $C_{\Delta r}(\tau)=\langle \Delta r(t)\:\Delta r(t+\tau) \rangle/\langle (\Delta r(t))^2 \rangle$ [Fig.~\ref{fig_corr_velocity_in_space}(d)] and the associated integral time scale $\tau_{\Delta r}\equiv \int_{0}^{\infty} C_{\Delta r}(\tau)d\tau$ [Fig.~\ref{fig_corr_velocity_in_space}(d, inset)]. As the droplet radius increases, $C_{\Delta r}(\tau)$ decays more rapidly with lag time $\tau$, thus $\tau_{\Delta r}$ decreases with increasing radius. This indicates that larger droplets not only take relatively smaller steps $\Delta r$ [Fig.~\ref{fig_energy_vorticity_velocity} (k)], they also tend to forget what step size $\Delta r$ they had taken in the past relatively quickly [Fig.~\ref{fig_corr_velocity_in_space} (d)]. However, $\tau_{\Delta r}$ is associated with trajectories and tells little how the droplet of a given size interacts with the active bath. To understand this we compute two-point spatial velocity correlation function $C_{vv}(r)=\langle {\boldsymbol{u}}(r)\cdot{\boldsymbol{u}}(0) \rangle/\langle u(0)^2 \rangle$ of the background velocity field of the active bath, and associated integral length scale, $\ell_v\equiv \int_{0}^{\infty} C_{vv}(r)dr$ (see Fig.~\ref{fig_corr_velocity_in_space} (e-f)). We find that as we decrease the radius of the droplet below  $\ell_v$, the droplet executes relatively larger $\Delta r$ [Fig.~\ref{fig_corr_velocity_in_space} (c) and Fig.~\ref{fig_energy_vorticity_velocity} (k)]. Conversely droplets with size larger than $\ell_v$ undergo negligible overall displacement in a given time. The variance of step size distribution drops as the size of the droplet is increased, and if $R > \ell_v$ the variance drops to $\approx 1$ [Fig.~\ref{fig_corr_velocity_in_space} (c)]. We propose that droplets with $R\ll\ell_v$ experience relatively larger drifts and droplets with $R\gg\ell_v$ move relatively smaller distances because in the latter case the velocities at any two opposite ends of the droplet are expected to be decorrelated -- resulting in negligible net advection of the interface. It is also to be noted that although the change in radius alters the integral time scale $\tau_{\Delta r}$ associated with the step autocorrelation function [Fig.~\ref{fig_corr_velocity_in_space} (d, inset)], it has negligible influence on $\ell_v$ [Fig.~\ref{fig_corr_velocity_in_space} (e, inset)] implying that suspending a droplet has negligible effect on the velocity characteristics of the background active bath.

Our observations suggest that the random forces due to the active bath on the droplet are correlated and depends on the droplet size. Therefore, we assume the following form for the random force with a memory which decays exponentially in time~\cite{wu2000particle,nguyen2021active},
\begin{equation}
\langle F(t)F(t+\tau)\rangle = \Lambda \exp [-\tau/\tau_F],\;\langle F(t)\rangle=0,
\label{eq_random_force}
\end{equation}     
where $\tau_F$ is a characteristic time scale, and $\Lambda\equiv D\gamma^2/\tau_{F}$, where $\gamma$ is the coefficient of friction from the Langevin equation $m \ddot{r}(t)= -{\gamma} \dot{r}(t) + F(t)$.  Mean square displacement $\langle r(t)^2\rangle$ can be obtained (see Supplementary for detailed derivations) and is depicted in Fig.~\ref{fig_corr_velocity_in_space} (c, inset). The crossover from a ballistic $\langle r(t)^2\rangle \sim t^2$ to diffusive $\langle r(t)^2\rangle \sim t$ is observed. In the limit $t\gg\tau_F$, the velocity autocorrelation is given as 
\begin{equation}
\langle v(0)v(t)\rangle= \langle v(0)^2\rangle \exp\left[-\frac{ \tau_F\Lambda}{m k_\mathrm{B}T} \:t\right],
\end{equation}
which gives the correlation time $\tau_{\Delta r} = \left[{m k_\mathrm{B}T}/\tau_F\Lambda\right]  \equiv \left[m k_\mathrm{B}T/D\gamma^2\right]$. 
As we increase $R$, the mass and inertial delay time increases, but is not the only factor behind the change in the dynamics as from simulations we see that $\tau_{\Delta r}$ decreases as we increase $R$ (and $m$) of the droplet. This is because in the limit $t\ll\tau_F$ the velocity autocorrelation function have the form~\cite{wu2000particle} $\langle v(t)v(0)\rangle= (2D/\tau_F)\exp\left[-t/\tau_F\right]$.
In this limit, the correlations in the random active force dominate over the inertial effect. Additionally, the value of $R$ relative to $\ell_v$ can have significant effect on the change in the crossover time. Smaller particles in active bath take too long to pass the crossover to diffusive MSD regime because  $\tau_{\Delta r}$ increases -- as shown through in Fig.~\ref{fig_corr_velocity_in_space} (c, inset).

Our results can be experimentally tested, for example using a setup of quasi two dimensional suspension of beads and bacteria on a soap film~\cite{wu2000particle} or inside thin fluid films~\cite{zhang2009swarming}. Experiments on self-diffusion of tagged bacterium in a dense swarms grown on agar plates~\cite{ariel2015swarming}, or of amoeboid cells~\cite{cherstvy2018non}, already confirm existence of superdiffusive regime while it is hard to reach normal diffusive regime. For example, multi-potent progenitor cells exhibit superdiffusive regime which can be persistent even for hours~\cite{partridge2022heterogeneous}. Here we have pin pointed how the size of the droplet and the properties of the active bath changes the spatial and temporal scales~\cite{wu2000particle,zhang2009swarming,ariel2015swarming,cherstvy2018non,partridge2022heterogeneous}. Our numerical study will also add to the better understanding of proposed analytical theories of diffusion of bodies in active systems~\cite{ortlieb2019statistics,maitra2020enhanced,granek2022anomalous,angelani2011effective,chen2007fluctuations,maes2020fluctuating,peng2016diffusion,speck2021vorticity}.

We acknowledge support from High Performance Computing facility at IISER Mohali. CS acknowledges support from the INSPIRE Faculty Fellowship of the Department of Science and Technology, India.

%

%
%

\onecolumngrid
\section*{Supplemental Material: Dynamics of a Passive Droplet in Active Turbulence}

\section*{Rescaling of governing equations and simulation parameters}
If we substitute the following reduced or rescaled variables/fields/properties/operators into the dimensional form of the governing equations 
\begin{align}\nonumber
	\rho'=\rho/\rho_o,\;
	t'=t/t_o,\;
	\partial_t'=t_o\partial_t,\;
	\boldsymbol{u}'=\boldsymbol{u}/u_o,\;
	\nabla'=x_o \nabla,\;
	p'=p/p_o,\;
	\eta'=\eta/\eta_o,\;
	\boldsymbol{E}'=\boldsymbol{E}/E_o,\;
	{\boldsymbol{\sigma}^\textrm{P}}'={\boldsymbol{\sigma}^\textrm{P}}/{\sigma}^\textrm{P}_o,\\
	\zeta'=\zeta/\zeta_o,\;
	\boldsymbol{Q}'=\boldsymbol{Q}/q,\;
	\mu'=\mu/\mu_o,\;
	{\boldsymbol{f}^\mathrm{I}}'=\boldsymbol{f}^\mathrm{I}/f^\mathrm{I}_o,\;
	\Gamma'=\Gamma/\Gamma_o,\;
	\boldsymbol{H}'=\boldsymbol{H}/H_o,\;
	\boldsymbol{S}'=\boldsymbol{S}/S_o,\;
	K'=K/K_{o},\;
	C'=C/C_{o},
\end{align}
and if we fix the following scales
\begin{align}
	x_o=\sqrt{\frac{K_o}{\Gamma_oC_o\eta_o}},\quad
	t_o=\frac{1}{\Gamma_o C_o},\quad
	\rho_o=\frac{\eta_o t_o}{x_o^2},\quad
	p_o=\frac{\eta_o}{t_o},\quad
	u_o=\frac{x_o}{t_o},\quad		
	S_o=\frac{u_o}{x_o},\quad
	H_o=C_o={\sigma}^\textrm{P}_o=\frac{K_o}{x_o^2},\quad
	q=1,
\end{align}
then the dimensional form of the governing equations can be rescaled into the following non-dimensional form
\begin{align}
	\mathrm{Re}\;\rho'\:(\partial'_t\boldsymbol{u}'+ \boldsymbol{u}'\cdot \nabla' \boldsymbol{u}')&= \nabla'\cdot(-p'\boldsymbol{I}+2\eta'\boldsymbol{E}'+{\boldsymbol{\sigma}^\textrm{P}}'-\mathrm{Re_a}\;\zeta'\boldsymbol{Q}')-\mathrm{Re_f}\;\mu'\boldsymbol{u}'+\mathrm{Re_\mathrm{I}}\:{\boldsymbol{f}^\textrm{I}}',\\
	\partial_t'\boldsymbol{Q}' + \boldsymbol{u}' \cdot \nabla' \boldsymbol{Q}'&= \boldsymbol{S}'+ \Gamma'\boldsymbol{H}'
	\\	
	\boldsymbol{E}'&=[\nabla'\boldsymbol{u}'+(\nabla'\boldsymbol{u}')^T]/2\\
	\boldsymbol{\Omega}'&=[\nabla'\boldsymbol{u}'-(\nabla'\boldsymbol{u}')^T]/2\\	
	\boldsymbol{S}'&=(\lambda\boldsymbol{E}'+\boldsymbol{\Omega}')\cdot(\boldsymbol{Q}'+\boldsymbol{I}/d) 
	+(\boldsymbol{Q}'+\boldsymbol{I}/d)\cdot(\lambda\boldsymbol{E}'-\boldsymbol{\Omega}')
	-2\lambda(\boldsymbol{Q}'+\boldsymbol{I}/d)(\boldsymbol{Q}':\nabla'\boldsymbol{u}')\\
	\boldsymbol{H}'&=K'\nabla'^2\boldsymbol{Q}'+C'\boldsymbol{Q}'/3+C'(\boldsymbol{Q}'\boldsymbol{Q}'-\boldsymbol{Q}':\boldsymbol{Q}'\;\boldsymbol{I}/d)-C'\boldsymbol{Q}'(\boldsymbol{Q}':\boldsymbol{Q}')\\
	{\boldsymbol{\sigma}^\textrm{P}}'&=2\lambda(\boldsymbol{Q}'+\boldsymbol{I}/d)(\boldsymbol{Q}':\boldsymbol{H}')
	-\lambda\boldsymbol{H}'\cdot(\boldsymbol{Q}'+\boldsymbol{I}/d) \nonumber\\
	&-\lambda(\boldsymbol{Q}'+\boldsymbol{I}/d)\cdot\boldsymbol{H}'
	+\boldsymbol{Q}'\cdot\boldsymbol{H}'-\boldsymbol{H}'\cdot\boldsymbol{Q}'
	-K'\partial_i'Q_{kl}'\partial_j'Q_{kl}'
\end{align}
where $'$ denotes a non-dimensional variable/field/property/operator, and 
\begin{align}
	\mathrm{Re}=\frac{\rho_o u_o x_o}{\eta_o},\quad
	\mathrm{Re_a}=\frac{t_o \zeta_o}{\eta_o},\quad
	\mathrm{Re_f}=\frac{\mu_o x_o^2}{\eta_o},\quad
	\mathrm{Re_I}=\frac{x_o t_o f_o^\mathrm{I}}{\eta_o},\quad
\end{align}
are non-dimensional groups signifying the ratio of inertial force to viscous force, active force to viscous force, friction force to viscous force, and interfacial tension force to viscous force, respectively. Implementation of the interfacial force $\boldsymbol{f}^\textrm{I}$ is described in detail in~\cite{tryggvason2011direct}. Numerical values of the non-dimensional parameters and non-dimensional properties in the active and passive phases are summarized in Table~\ref{table1}. In addition to fluid properties in the active and passive phases, the table also summarizes other simulation parameters namely the system size, grid size, time step, and droplet radius.
The grid and time steps are taken such that the advective as well as viscous time step conditions are well satisfied, namely~\cite{tryggvason2011direct}
\begin{equation}
	\Delta t < C_1 \frac{\Delta}{u_\mathrm{max}},  \: \Delta t < C_2\frac{2\eta_\mathrm{min}}{\rho_\mathrm{max} u_\mathrm{max}^2}, \: \Delta t < C_3 \frac{\rho_\mathrm{min}\Delta^2}{4\eta_\mathrm{max}},
\end{equation}
where $\Delta$ is the grid step $L_x/M=L_y/N$, and subscripts $\mathrm{max, min}$ denote maximum and minimum values of the repective properties from the two phases. For simulation to remain stable, the factors $C_1, C_2, C_3$ have to be $<1$, and to be conservative, we choose time and grid size combination such that these conditions are always satisfied.

\begin{table}[t!]
	\begin{ruledtabular}
		\begin{tabular}{lcc}
			\textrm{Non-dimensional parameter/property}& 
			\textrm{Phase 1: active nematic fluid}&
			\multicolumn{1}{c}{\textrm{Phase 2: passive nematic droplet}}\\
			\colrule			
			\\			
			$\mathrm{Re}$& \multicolumn{2}{c}{$0.1$}\\				
			$\mathrm{Re_a}$& \multicolumn{2}{c}{$0.5$}\\
			$\mathrm{Re_f}$& \multicolumn{2}{c}{$0.00075$}\\
			$\mathrm{Re_I}$& \multicolumn{2}{c}{$7.0$}\\			
			\\
			System size $L_x\times L_y$&  \multicolumn{2}{c}{$200\times200,400\times400$}\\
			Grid size $M\times N$&  \multicolumn{2}{c}{$256\times256,512\times512$}\\
			Droplet radius $R$ &  \multicolumn{2}{c}{varied from $15$ to $45$}\\
			Time step $\Delta t$&  \multicolumn{2}{c}{$1.0\times10^{-3}$}\\			
			\\			
			Density $\rho'$ & $ \rho_1'=1.0$ & $\rho_2'=2.0$ \\
			Viscosity $\eta'$ &  $\eta_1'=1.0$ & $\eta_2'=2.0$\\
			Activity $\zeta'$ &  $\zeta_1'=1.0$ & $\zeta_2'=0.0$\\
			Friction $\mu'$ &  $\mu_1'=1.0$ & $\mu_2'=1.0$\\	
			Alignment $\lambda$ &  $\lambda_1=0.8$ & $\lambda_2=0.8$\\			
			Elastic constant $K'$ &  $K_{1}'=0.25$ & $K_{2}'=0.25$\\
			Material constant $C'$ &  $C_{1}'=0.4$ & $C_{2}'=0.4$\\	
			Relaxation $\Gamma'$ &  $\Gamma_1'=1.0$ & $\Gamma_2'=1.0$\\

		\end{tabular}
	\end{ruledtabular}
	\caption{\label{table1}%
		Numerical values of parameters and fluid properties adopted in the simulations. The above set of parameters are close to, for instance, the studies carried out by~\cite{thampi2015intrinsic,koch2021role}.
	}
\end{table}
\section*{Mean square displacement and velocity autocorrelation function}
To incorporate the memory effect in the stochastic force in the Langevin equation, let us start from an exponentially correlated random force
\begin{equation}
	\langle F(t)F(t+\tau)\rangle = \Lambda \exp [-\tau/\tau_F],\;\langle F(t)\rangle=0,
	\label{eq_random_force}
\end{equation}     
where $\tau_F$ is the time scale which tells us the extent of the force memory, and $\Lambda\equiv D\gamma^2/\tau_{F}$ with $D$ and $\gamma$ being diffusion constant and coefficient of friction respectively. For simplicity of the argument, only a component of the force is considered and the formulation can readily be extended to higher dimensions. The equation of motion for a passive particle in such an active bath can be written as
\begin{equation}
	\frac{d^2}{dt^2}r(t)= -\frac{\gamma}{m} \frac{d}{dt} r(t) + \frac{1}{m}F(t)= -\Gamma \frac{d}{dt} r(t) + \frac{1}{m}F(t),\quad \Gamma\equiv\gamma/m
	\label{eq_langevin}
\end{equation}
Multiplying the equation of motion with $x$ and $v\equiv\frac{d}{dt}x$ respectively, and taking the ensemble average, after few algebraic manipulations we obtain the equations for the mean square displacement (MSD) and the mean squared velocity, written as
\begin{align}
	\frac{d^2}{dt^2} \langle r(t)^2\rangle + \Gamma \frac{d}{dt} \langle r(t)^2\rangle &= 2 \langle v(t)^2\rangle + \frac{2}{m}\langle r(t)F(t)\rangle,\label{eq_langevin_msd}\\
	\frac{d}{dt} \langle v(t)^2\rangle + 2\Gamma \langle v(t)^2\rangle &= \frac{2}{m}\langle v(t)F(t)\rangle.
	\label{eq_langevin_energy}
\end{align}
While the correlation function $\langle r(t)F(t)\rangle=0$ (force on the droplet by the active bath is independent of the position of the droplet), the correlation function $\langle v(t)F(t)\rangle$ is in general non-zero and carries information about the balance between the energy supplied by the active bath to the droplet and the energy dissipated by the droplet due to the drag offered by the bath -- the fluctuation-dissipation mechanism. This correlation function can be obtained by multiplying $F(t)$ on both sides of the trivial identity $\int_{t-s}^{t}\frac{d}{dt'}v(t')dt'=v(t) - v(t-s)$, and taking the ensemble average
\begin{align}
	\int_{t-s}^{t}\langle\frac{d}{dt'}v(t')F(t)\rangle dt'=\langle v(t)F(t) \rangle - \langle v(t-s)F(t)\rangle.
\end{align}
The second term on right hand side is zero (velocity at some earlier time has to be independent of the force at a later time), and we input $\frac{d}{dt'}v(t')$ from the equation of motion to obtain  
\begin{align}
	\langle v(t)F(t) \rangle = -\Gamma \int_{t-s}^{t}\langle v(t')F(t)\rangle dt'+\frac{1}{m}\int_{t-s}^{t}\langle F(t')F(t)\rangle dt'.
\end{align}
The first term on right hand side is again zero (velocity at some earlier time has to be independent of the force at a later time). Taking the limit $s\rightarrow\infty$, and using exponentially correlated force from Eq.~\ref{eq_random_force}, we get
\begin{align}
	\langle v(t)F(t) \rangle = \frac{1}{m}\int_{0}^{\infty}  \Lambda \exp [-t'/\tau_F] dt' = \frac{ \tau_F\Lambda}{m}.
	\label{eq_velocity_force_corr}
\end{align}
Using Eq.~\ref{eq_velocity_force_corr} in Eq.~\ref{eq_langevin_energy}, and assuming statistically stationary state for $\langle v(t)^2\rangle \equiv k_\mathrm{B}T/m$, we find that
\begin{align}
	\Gamma 
	= \frac{1}{2\langle v(t)^2\rangle}\frac{2}{m}\langle v(t)F(t) \rangle
	= \frac{ \tau_F\Lambda}{m k_\mathrm{B}T}.
\end{align}
Here $T$ is an effective temperature of the active bath. Using this in Eq.~\ref{eq_langevin_msd}, we find the equation for the MSD for an exponentially correlated noise
\begin{align}
	\frac{d^2}{dt^2} \langle r(t)^2\rangle + \frac{ \tau_F\Lambda}{m k_\mathrm{B}T} \frac{d}{dt} \langle r(t)^2\rangle = \frac{2k_\mathrm{B}T}{m}
	\label{eq_msd}
\end{align}
So we find that the time scale $\tau_F$ of the exponentially correlated random force changes the crossover time of transition from ballistic to diffusive regime. It is important to note that there appears no such time scale in equation for MSD in case of delta correlated random force. Force time scale $\tau_F$ can be related to velocity correlation time scale $\tau_{\Delta r}$ as follows. Let us multiply the Langevin equation with $v(0)$ and then take the ensemble average, we get
\begin{equation}
	\frac{d}{dt}\langle v(0)v(t)\rangle= -\Gamma \langle v(0)v(t)\rangle + \frac{1}{m}\langle v(0)F(t)\rangle.
	\label{eq_langevin}
\end{equation}
The last term is zero by the same argument that velocity at some earlier time has to be independent of the force at a later time. Thus we have 
\begin{equation}
	\langle v(0)v(t)\rangle= \langle v(0)^2\rangle \exp\left[-\Gamma\:t\right] = \langle v(0)^2\rangle \exp\left[-\frac{ \tau_F\Lambda}{m k_\mathrm{B}T} \:t\right] .
\end{equation}
This provides us an estimate of the correlation time  $\tau_{\Delta r}$ associated with the velocity autocorrelation function, {\it i.e.}
\begin{equation}
	\tau_{\Delta r}=\frac{m k_\mathrm{B}T}{ \tau_F\Lambda}=\frac{m k_\mathrm{B}T}{ D\gamma^2}.
\end{equation}
as $\Lambda\equiv D\gamma^2/\tau_{F}$. The above result is valid in the limit $t\gg\tau_F$, and it can be proved that in the limit $t\ll\tau_F$
\begin{equation}
	\langle v(0)v(t)\rangle = \langle v(0)^2\rangle \exp\left[-\tau_F^{-1} \:t\right] .
\end{equation}.

\begin{figure}[t!]
	\centering
	\centering
	\includegraphics[width=1.0\linewidth]{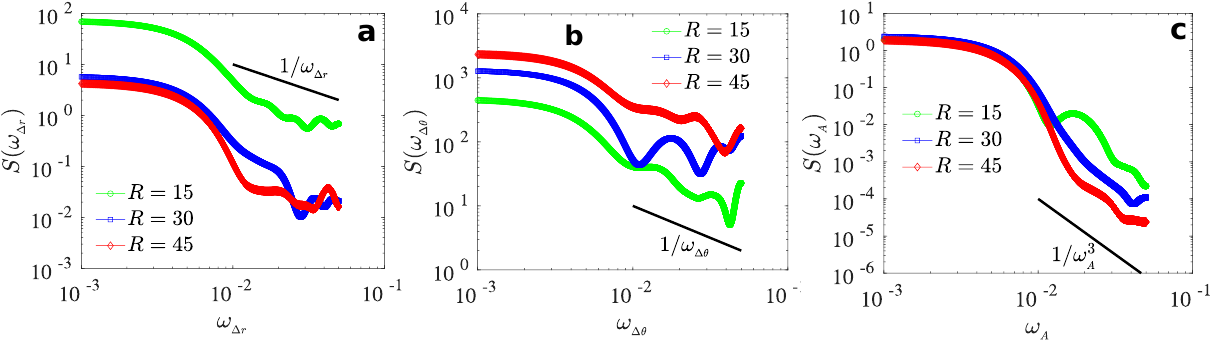}
	\caption{(a,b,c) Power spectrum of the time series of (a) translational step size $\Delta r$, (b) change in direction of motion $\Delta \theta$, and (c) droplet aspect ratio or deformation parameter $A$, respectively. Panels (a) implies that two consecutive droplet position increments are not independent and thus the case is different from standard Brownian increments as Brownian increments are known to be independent. Similarly panel (b) imlies that two consecutive $\Delta \theta$ are not independent.}
	\label{fig_spectral}
\end{figure}
\section*{Spectral analysis and discussion}
Figure~\ref{fig_spectral} show the power spectrum of signals: (a) translational step size $\Delta r(t)$, (b) change in direction of motion $\Delta \theta(t)$, and (c) droplet aspect ratio $A(t)$, respectively. The indication from the analysis is that the translational step and the angular change in the direction of motion signals exhibit nearly $1/\omega$ or pink noise. The spectrum of the aspect ratio or the deformation parameter $A(t)$, however, exhibits nearly $1/\omega^3$ noise. Although its known that $1/\omega$ noise is present in many natural, man made, and scientific systems~\cite{weissman19881,gisiger2001scale,gilden19951,hooge1981experimental}, its underpinnings are debated over decades~\cite{bak59self,bak1988self,bedard2006does}. Does above $1/\omega$ noise has to do with self-organized states in our system? In addition, $1/\omega^3$ noise in the deformation or the droplet aspect ratio signal indicates to us that deformations somehow have distinct dynamics than the motion increments. This type of noise is usually observed in oscillators implying that the droplet deformation has some oscillatory feature. Figure~\ref{fig_spectral} (a) implies that two consecutive droplet position increments are not independent and thus the case is different from standard Brownian increments as Brownian increments are independent. Similarly Fig.~\ref{fig_spectral} (b) imlies that two consecutive $\Delta \theta$ are not independent.

\end{document}